\documentclass[fleqn,usenatbib]{mnras}

\usepackage{newtxtext,newtxmath}
\usepackage[T1]{fontenc}
\usepackage{ae,aecompl}

\usepackage{graphicx}	% Including figure files
\usepackage{amsmath}	% Advanced maths commands
\usepackage{amssymb}	% Extra maths symbols
\usepackage{float}
\usepackage{subcaption}

\usepackage{bold-extra}
\usepackage{ulem}

\usepackage{xspace}

\newcommand{\Msun}{M_{\sun}}
\newcommand{\hkpc}{h^{-1} {\rm kpc}}
\newcommand{\hMpc}{h^{-1} {\rm Mpc}}
\newcommand{\hMsun}{h^{-1} \Msun}

\newcommand{\Mbh}{M_{\bullet}}
\newcommand{\Mstar}{M_{\star}}
\newcommand{\Mhalo}{M_{\rm halo}}

\newcommand{\MbhSeed}{\Mbh^{\rm seed}}
\newcommand{\MfofSeed}{M_{\rm fof}^{\rm seed}}

\newcommand{\BT}{\textsc{BlueTides}\xspace}

\newcommand{\code}[1]{\textsc{#1}\xspace} %MNRAS style is small caps
\newcommand{\acknowledgeSoftware}[1]{\code{#1} \citep{#1}\xspace}

\title[Black hole growth in cosmological simulations]{The early growth of supermassive black holes in cosmological hydrodynamic simulations with constrained Gaussian realizations}

\author[K.-W. Huang et al.]{Kuan-Wei Huang$^{1}$\thanks{E-mail: kuanweih@andrew.cmu.edu},
Yueying Ni$^{1}$,
Yu Feng$^{2}$,
Tiziana Di Matteo$^{1,3}$
\\
$^{1}$McWilliams Center for Cosmology, Dept. of Physics, Carnegie Mellon University, Pittsburgh, PA, 15213, USA\\
$^{2}$Berkeley Center for Cosmological Physics, University of California at Berkeley, Berkeley, CA, 94720, USA\\
$^{3}$School of Physics, The University of Melbourne, VIC 3010, Australia
}

\date{Accepted XXX. Received YYY; in original form ZZZ}

\pubyear{2019}

\begin{document}
\label{firstpage}
\pagerange{\pageref{firstpage}--\pageref{lastpage}}
\maketitle

\begin{abstract}
The paper examines the early growth of supermassive black holes (SMBHs) in cosmological hydrodynamic simulations with different BH seeding scenarios.
Employing the constrained Gaussian realization, we reconstruct the initial conditions in the large-volume \BT simulation and run them to $z = 6$ to cross-validate that the method reproduces the first quasars and their environments.
Our constrained simulations in a volume of $(15 \, \hMpc)^3$ successfully recover the evolution of large-scale structure and the stellar and BH masses in the vicinity of a $\sim 10^{12} \, \Msun$ halo which we identified in \BT at $z \sim 7$ hosting a $\sim 10^9 \, \Msun$ SMBH.
Among our constrained simulations, only the ones with a low-tidal field and high-density peak in the initial conditions induce the fastest BH growth required to explain the $z > 6$ quasars.
We run two sets of simulations with different BH seed masses of $5 \times 10^3$, $5 \times 10^4$, and $5 \times 10^5 \, \hMsun$, (a) with the same ratio of halo to BH seed mass and (b) with the same halo threshold mass.
At $z = 6$, all the SMBHs converge in mass to $\sim 10^9 \, \Msun$ except for the one with the smallest seed in (b) undergoing critical BH growth and reaching $10^8$ -- $10^9 \, \Msun$, albeit with most of the growth in (b) delayed compared to set (a).
The finding of eight BH mergers in the small-seed scenario (four with masses $10^4$ -- $10^6 \, \Msun$ at $z > 12$), six in the intermediate-seed scenario, and zero in the large-seed scenario suggests that the vast BHs in the small-seed scenario merge frequently during the early phases of the growth of SMBHs.
The increased BH merger rate for the low-mass BH seed and halo threshold scenario provides an exciting prospect for discriminating BH formation mechanisms with the advent of multi-messenger astrophysics and next-generation gravitational wave facilities.
\end{abstract}


%%%%%%%%%%%%%%%%%%%%%%%%%%%%%%%%%%%%%%%%%%%%%%%%%%

%%%%%%%%%%%%%%%%% BODY OF PAPER %%%%%%%%%%%%%%%%%%
\section{Introduction}
\label{sec:Intro}

The formation of the first supermassive black holes (SMBHs) remains challenging in our standard paradigm of structure formation. 
SMBHs, as massive as those in galaxies today, are known to exist in the early universe, even up to $z \sim 7.5$. 
Luminous, extremely rare, quasars at $z \sim 6$ were initially discovered in the Sloan Digital Sky Survey \citep{Fan2006, Jiang2009} and, until recently, the highest redshift quasar known \citep{Wu2015} at $z = 7.09$ \citep{Mortlock2011} has been surpassed by the discovery of a bright quasar at $z = 7.54$ \citep{Banados2018}, which is currently the record holder for known high redshift quasars. 
A further sample of two $z > 7$ has also been recently discovered \citep{Yang2018}. 
The presence of luminous quasars observed within the first billion years of the Universe highlights that the BH seeds for the SMBH population must have assembled at the cosmic dawn, concurrently with the time of the formation of the first stars or galaxies.

However, the precise SMBH seed formation mechanism remains unknown, nor is it clear if there is only one seed formation channel at play over the entire SMBH seed mass spectrum of models.
The current scenarios suggest that seed BHs are (a) remnants of the first generation of stars (PopIII) \citep[e.g.][]{MadauRees2001, Abel2002, Johnsonbromm2007} or (b) direct gas collapse within the first massive halos \citep[e.g.][]{Lodato2006, Begelman2006, ReganHaehnelt2009, Ferrara2014, Latif2013} or (c) runaway collapse of dense nuclear star clusters \citep[e.g.][]{Begelman1978, Devecchi2009, Yajima2016, Katz2015}. 
The seed BHs then range in mass from a few hundred for (a) to $10^5 \Msun$ for (b) and (c).

In large-volume cosmological simulations, a common and widely used sub-grid model for SMBHs and active galactic nuclei (AGN) feedback has been proposed in \citet{DiMatteo2005}.
Since the SMBH seed formation process is not resolved by cosmological simulations \cite[see][for a review]{ReganHaehnelt2009}, it is assumed that every halo above a certain threshold mass hosts a central BH seed. 
Halos are selected for seeding by regularly running the 'Friends-of-Friends' (FoF) halo finder on the dark matter distribution.  
The BH seed mass ($\MbhSeed$) and the threshold halo mass ($\MfofSeed$) are the parameters in simulations. 
Although this is an ad-hoc seeding procedure, the initial seed BH mass subsequently grows in these simulations via mergers and accretion.
Many simulations have adopted this or a similar scenario and gotten good agreements with observations such as the \textsc{MassiveBlack} simulation \citep{2012ApJ...745L..29D}, the \textsc{Illustris} simulation \citep{Vogelsberger2014}, the Evolution and Assembly of GaLaxies and their Environment (\textsc{EAGLE}) suite of SPH simulation \citep{Schaye2015}, the \textsc{MassiveBlack II} simulation \citep{Khandai2015}, and the \BT simulation \citep{Feng2016}. 
Some recent studies have implemented different, physically motivated approaches where the BH seeding is based on gas properties such as \citet{Bellovary_2011,2017MNRAS.468.3935H,Tremmel2017}. 
However, it is worth noting that these models were adopted in much smaller volume simulations than, for example, \textsc{MassiveBlack II} or \BT. 
From those, it is not possible to validate the basic statistical properties of the BH population (e.g. luminosity functions or mass functions) against currently observed samples.

Taking \BT, a large-volume and high-resolution cosmological hydrodynamic simulation with $2 \times 7040^3$ particles in a box of $400 \, \hMpc$ on a side, as an example, SMBHs are modeled as follows. 
For each FoF halo with a mass above $\MfofSeed = 5 \times 10^{10} \, \hMsun$, a SMBH is seeded with an initial seed mass $\MbhSeed = 5 \times 10^5 \, \hMsun$ at the position of the local minimum potential if there is no SMBH in that halo. 
After being seeded, gas accretion proceeds according to \citet{Bondi1952} while the BH accretion rate is limited to two times the Eddington rate. 
When SMBHs are accreting, we assume that some fraction of the radiated luminosity can couple thermally and isotropically to surrounding gas in the form of feedback energy \citep{Springel2005, DiMatteo2005}.

Adopting this SMBH model and appropriate sub-grid physics for the galaxy formation modeling, the \BT simulation has predicted various quantities in good agreements with current observational constraints in the high-$z$ universe such as UV luminosity functions \citep{Feng2016, Waters2016a, Waters2016b, Wilkins2017}, the first galaxies and the most massive quasars \citep{Feng2015, DiMatteo2017, Tenneti2018}, the Lyman continuum photon production efficiency \citep{Wilkins2016, Wilkins2017}, galaxy stellar mass functions \citep{Wilkins2018}, angular clustering amplitude \citep{Bhowmick2017}, BH-galaxy scaling relations \citep{Huang2018}, and gas outflows from the $z = 7.54$ quasar \citep{Ni2018}.
Important for our work here, \BT, with its large volume and appropriate resolution, is currently the only cosmological hydrodynamic simulation that makes direct contact with the rare, first quasar population at $z > 7$.

However, an essential question for the SMBH sub-grid model is how different parameters (e.g. $\MbhSeed$ or $\MfofSeed$) may affect the growth of SMBHs and the local environment in cosmological simulations.
Changing the BH seed mass and re-running such a large-volume simulation multiple times is completely prohibitive because it is computationally expensive even on the largest current national facilities. 
To reduce the demand on computational resources, a common method is to run a "zoom-in re-simulation" with a higher resolution or different physical parameters from a certain region selected from a large-volume lower-resolution simulation. 
This allows people to focus on a specific environment numerically and has been applied to study SMBHs in simulations for various purposes \citep[e.g.][]{Li2007ApJ...665..187L, Sijacki2009MNRAS.400..100S, Hopkins2010MNRAS.407.1529H, Bournaud2011ApJ...741L..33B, Romano-Diaz2011ApJ...736...66R, Bellovary2013ApJ...779..136B, Dubois2013MNRAS.428.2885D, Angles-Alcazar2014ApJ...782...84A, Costa2014MNRAS.439.2146C, Yu2014MNRAS.440.1865F}.

In this paper, we combine the technique of constrained Gaussian realization and cosmological hydrodynamic simulations to reduce the demand on computational resources. 
\citet{Hoffman1991} first introduced an optimal solution to the problem of the construction of constrained realizations of Gaussian fields by demonstrating how the algorithm generates constrained fields with a simple single-density peak.
Later on, \citet{vandeWeygaert1996} in addition developed an algorithm to set up initial Gaussian random density and velocity fields containing multiple constraints of arbitrary amplitudes and positions. 
Integrating the algorithm to cosmological hydrodynamic simulations has arisen in the past few years to explore dark matter halos and galaxy formation \citep{2016MNRAS.455..974R, 2016MNRAS.463.4068P, 2017MNRAS.465..547P}.

With the constrained Gaussian realization, we can constrain the initial density field by adding a desirable height of a density peak when generating the initial condition such that a more massive halo can still form in a relatively small box compared to those large-volume ($\sim {\rm Gpc}$ per side) cosmological simulations (with uniform/unconstrained initial condition) . 
For instance, we can grow a halo with a mass $\sim 10^{12} \, \Msun$ in a box of $15 \, \hMpc$ on a side at $z=8$, whose mass is similar to the one hosting the most massive BH in \BT simulation ($400 \, \hMpc$ on a side) under the same resolution. This reduces the computational demand by a factor of $(400/15)^3 \sim 20000$. 
This approach is a more general way to study the growth of SMBH compared to the zoom-in re-simulation method because the goal of the latter is to exactly study a particular object/region (for example, a particular halo). 
However, our approach is aiming to study characteristic environments by creating one or more different realizations but with similar properties such as halo mass or tidal field to the object/region of our interest (which we extract, for comparison, in the uniform large volume simulations with the exact physics).

As we shall further demonstrate, besides the density constraint, we need another condition related to the ICs that induces the fastest growth for SMBHs in cosmological simulations. 
This is expected, as the observed population of quasar-like SMBHs at high redshifts is even rarer than the massive halos. 
For example, there is only one SMBH with a mass above $10^8 \, \Msun$ in a halo with a mass $\sim 10^{12} \, \Msun$ while there are $> 50$ halos more massive than that halo at $z = 8$. 
An environmental property, also related to the ICs that has been found to be relevant to induce fast BH growth is the local tidal field strength \citep{DiMatteo2017}. 
In particular, using the large volume \BT simulations, \citet{DiMatteo2017} has shown that isolated regions of low tidal fields are key to the fast growth of the firs SMBHs. 
As a consequence, we also choose the realization with a lower tidal field around the local environment where the halo forms, which indeed helps more massive SMBHs grow in simulations (see Section~\ref{subsec:t1}).

After significantly decreasing the demand on computational resources with the constrained Gaussian realizations and a lower tidal field realization, we are finally able to examine how sensitive the SMBH growth is to the BH seed mass in the sub-grid model by running multiple cosmological simulations. 
According to the different hypotheses of the BH formation scenario, the BH seed mass has been suggested to range from $10^2$ to $10^6 \, \hMsun$ \citep{Haehnelt1993, Loeb1994, Eisenstein1995, Bromm2003, Koushiappas2004, Begelman2006, Lodato2006, Zhang2008, Volonteri2010, Latif2013, Schleicher2013, Ferrara2014}. 
Therefore we focus on the three different seed masses: $5 \times 10^3$, $5 \times 10^4$, and $5 \times 10^5 \, \hMsun$, with the same ratio of halo to BH seed mass and with the same halo threshold mass in this paper.

We organize the paper as the following.  
Section~\ref{sec:method} describes the constrained Gaussian realizations, compares the constrained and unconstrained initial conditions and simulations, and discusses the effect of different tidal fields on the local environment of SMBHs. 
Section~\ref{sec:Results} demonstrates the results of the early growth of SMBHs and their hosts in the simulations with different BH seeding scenarios. 
Section~\ref{sec:Conclusions} concludes the paper.

\section{methodology}
\label{sec:method}

% --------------------------------------------------
\subsection{Constrained initial conditions}
\label{sec:ICs}
% --------------------------------------------------

% Fig ----------------------------------------------
\begin{figure*}
    \includegraphics[width=2\columnwidth]{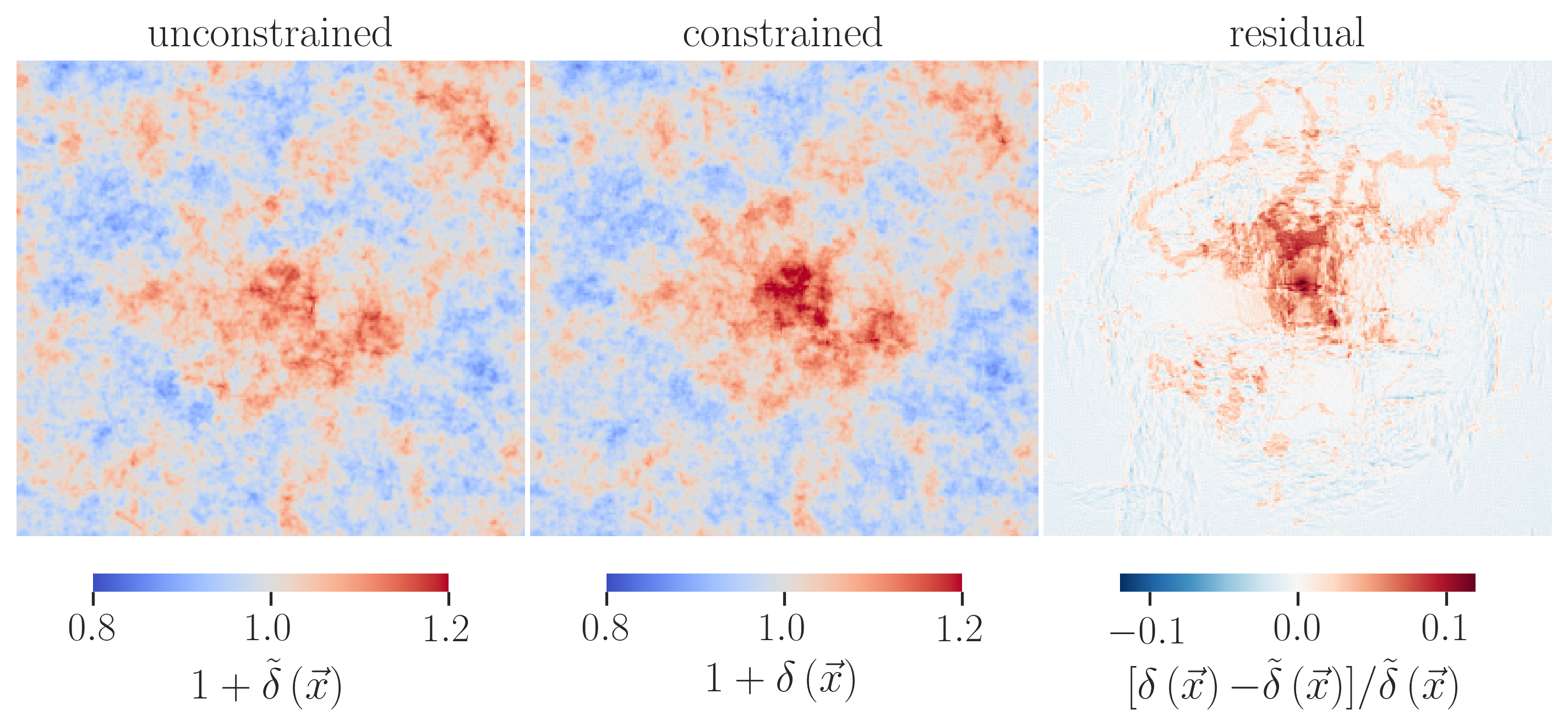}
    \caption{Slices of density fields of the initial conditions with the same realization number. \textbf{Left}: without any constraints. \textbf{Middle}: with a constrained density peak at the densest region of the original field (also the center of the panel). The boxes are $15 \, \hMpc$ per side with a thickness of $5 \, \hMpc$. \textbf{Right}: the residual of the constrained and unconstrained density fields.}
    \label{fig:ic_illustration}
\end{figure*}
% Fig ----------------------------------------------

% --------------------------------------------------
The first quasars are extremely rare and hence the massive halos hosting these first SMBHs also have to be commensurately rare. Traditionally one needs extremely large-volume simulations to simulate such objects.
For example, there is only one SMBH with a mass above $10^8 \, \Msun$ in a halo with a mass of $\sim 10^{12} \, \Msun$ at $z = 8$ in the \BT simulation with $400 \, \hMpc$ on a side of the cube.
Using simulations of such scale to study the effect on BH growth due to subgrid prescriptions is prohibitively expensive.
In this work, we work with much smaller simulation boxes ($15 \, \hMpc$), but we add constraints to the initial condition (linear field) to ensure the existence of extreme density peaks.
With high-density peaks in the linear field, we can guarantee the formation of massive halos at later times even in a smaller simulation box.
These halos are then used to study the early growth of SMBHs in the massive halos in cosmological hydrodynamic simulations with different BH seeding parameters.

According to the linear growth theory, to generate a $10^{12} \, \Msun$ halo at z=8, we need a $5 \, \sigma$ peak in the underlying over-density field.
To achieve that, we use \textsc{FastPM}, a particle mesh based quasi N-body solver \citep{Feng2016fastpm}, to generate initial conditions with constrained Gaussian realizations for the simulations in the paper for the first time.
Thanks to the contribution of \citet{Aslanyan2016unpublished}, \textsc{FastPM} is capable of producing constrained Gaussian density fields to the initial condition, which we will describe the basic idea and the implementation in the following paragraphs.

First introduced by \citet{Hoffman1991}, the constrained realization technique was then explained in a lot more detail in \citet{vandeWeygaert1996} which we refer readers to.
The goal is to construct a field $f(\mathbf x)$ subject to a set of $M$ constraints:
\begin{equation}
    \Gamma = \{ C_i \equiv C_i [f; \mathbf x_i] = c_i; \, i = 1,...,M \}
\end{equation}
where the constraint $C_i$ can be viewed as a functional of $f(\mathbf x)$ field (here in our specific case, the overdensity field) to have the specific value $c_i$ at the position $\mathbf x_i$.

To obtain a field $f(\mathbf x)$ satisfying the constraint $\Gamma$, one can start with a random, unconstrained Gaussian realization $\tilde{f}(\mathbf x)$ and impose on that an "ensemble mean field" $\bar{f}(\mathbf x)$ corresponding to the desired constraint $\Gamma$.
More specifically, the ensemble mean field $\bar{f}(\mathbf x)$ can be written in the form of:
\begin{equation}
\label{equation:Ensemble_mean}
    \bar{f}(\mathbf x) = \langle f(\mathbf x) \, | \, \Gamma \rangle = \xi_i(\mathbf x) \, \xi^{-1}_{ij} \, c_j
\end{equation}
where $\xi_i(\mathbf x) = \langle f(\mathbf x) \, C_i \rangle$ is the cross-correlation between the $f(\mathbf x)$ field and the $i^{\rm th}$ constraint $C_i$, and $\xi^{-1}_{ij}$ is the $\left( i, j \right)$ element of the inverse of the constraint covariance matrix $\langle C_i \, C_j \rangle$.
Note that the summation over repeated indices is used.
The ensemble mean field $\bar{f}(\mathbf x)$ can be interpreted as the "most likely" field subject to the set of constraints $\Gamma$.

We further introduce the "residual field" $F(\mathbf x) \equiv f(\mathbf x) - \bar{f}(\mathbf x)$ as the difference between an arbitrary Gaussian realization $f(\mathbf x)$ satisfying the constraint set $\Gamma$ and the ensemble mean field $\bar{f}(\mathbf x)$.
The crucial idea of the constrained realization construction method is based on the fact that, the complete probability distribution $\mathcal{P} [F \, | \, \Gamma]$ of the residual field $F(\mathbf x)$ is independent of numerical values $c_i$ of the constraints in $\Gamma$ \citep[c.f.][for the detailed derivations]{Hoffman1991,vandeWeygaert1996}.
That is, for any $\Gamma_p$ and $\Gamma_q$ where $p \neq q$,
\begin{equation}
    \mathcal{P} [F \, | \, \Gamma_p] = \mathcal{P} [F \, | \, \Gamma_q]
\end{equation}
Therefore, we can construct the desired realization under a constraint set $\Gamma$ by properly sampling the residual field $F(\mathbf x)$ from a random, unconstrained realization $\tilde{f} (\mathbf x)$ and then add that $F(\mathbf x)$ to the ensemble field $\bar{f}(\mathbf x)$ corresponding to $\Gamma$.
The formalism can be written as
\begin{equation}
\label{equation:construct_fx}
\begin{split}
    f(\mathbf x)
        & = F(\mathbf x) + \bar{f}(\mathbf x) \\
        &= \left( \tilde{f} (\mathbf x) - \xi_i(\mathbf x) \, \xi^{-1}_{ij} \, \tilde{c}_j \right)  + \xi_i(\mathbf x) \, \xi^{-1}_{ij} \, c_j \\
        &= \tilde{f} (\mathbf x)  + \xi_i(\mathbf x) \, \xi^{-1}_{ij} \, \left( c_j - \tilde{c}_j \right)
\end{split}
\end{equation}
In other words, we treat the original $\tilde{f}(\mathbf x)$ as a field subject to a constraint set $\tilde{\Gamma}$ with $\tilde{c}_j = C_j [\tilde{f}; \mathbf x_j]$ where $\tilde{c}_j$ is the original value of the unconstrained field.
We then have the ensemble mean field corresponding to $\tilde{\Gamma}$ as $\tilde{\bar{f}}(\mathbf x) = \xi_i(\mathbf x) \, \xi^{-1}_{ij} \, \tilde{c}_j$.
Getting the residual field $F(\mathbf x)$ from a random unconstrained realization by $\tilde{f}(\mathbf x) - \tilde{\bar{f}}(\mathbf x)$, we then add $F(\mathbf x)$ to $\bar{f}(\mathbf x)$ to obtain the field $f(\mathbf x)$ satisfying the constraint $\Gamma$.
It is well established in \citet{vandeWeygaert1996} that the $f(\mathbf x)$ field constructed in this way is a properly sampled realization subject to the desired constraint $\Gamma$.
This is what we implemented in our code.

We note that, however, one limitation in our implementation is that the super-sampling variance (DC mode) is missing. Super-sampling variance is the effect of the coupling to modes at scales larger than the box size \citep{Li2014}.
In our simulation, we assume that the overdensity of the whole simulation box to be zero, i.e., the DC mode is zero. The DC mode can be incorporated by the so-called separate universe technique that absorbs the overdensity of the simulation volume into a modified cosmology.
\citep[see, e.g.,][for more details]{Sirko2005,Gnedin2011,Wagner2015,Li2014,Li2018}.
However, we estimate that with our simulation box size ($15 \, \hMpc$ per side) at these high redshifts ($z > 6$), this effect accounts only about 10 percent.

Here we demonstrate the constrained realization generated via \textsc{FastPM} with a single $5 \, \sigma$ density peak.
Figure~\ref{fig:ic_illustration} shows examples of the density field with and without a constrained density peak and its associated residual map in a dense region at the center in the domain with a box size of $15 \, \hMpc$.
As expected, we find that the density increases in the region where we put the constraint without changing the overall pattern of the density field.
According to the residual map, 0.04 percent of the pixels have greater than 10 percent residuals; 2 percent of the pixels exceed 5 percent residuals; none of the pixels have residuals less than $-5$ percent.
% --------------------------------------------------

% --------------------------------------------------
\subsection{Simulation setup}
\label{sec:setup}
% --------------------------------------------------

% Table --------------------------------------------
\begin{table}
    \centering
    \caption{Parameters adopted in our simulations.}
    \label{tab:suites}
    \begin{tabular}{llllll}
        \hline
        $h$        & 0.697          & $\Omega_{\rm matter}$  & 0.2814           & $M_{\rm DM}$  & $1.7 \times 10^7 \, \Msun$ \\
        $\sigma_8$ & 0.820          & $\Omega_{\rm baryon}$  & 0.0464           & $M_{\rm gas}$ & $3.4 \times 10^6 \, \Msun$ \\
        $n_s$      & 0.971          & $\Omega_{\rm \Lambda}$ & 0.7186           & $\Mstar$      & $8.4 \times 10^5 \, \Msun$ \\
        $\epsilon$ & $1.5 \, \hkpc$ & $N_{\rm particle}$     & $2 \times 264^3$ & $L_{\rm box}$ & $15 \, \hMpc$ \\
        \hline
    \end{tabular}
\end{table}
% Table --------------------------------------------

% Fig ----------------------------------------------
\begin{figure}
    \includegraphics[width=\columnwidth]{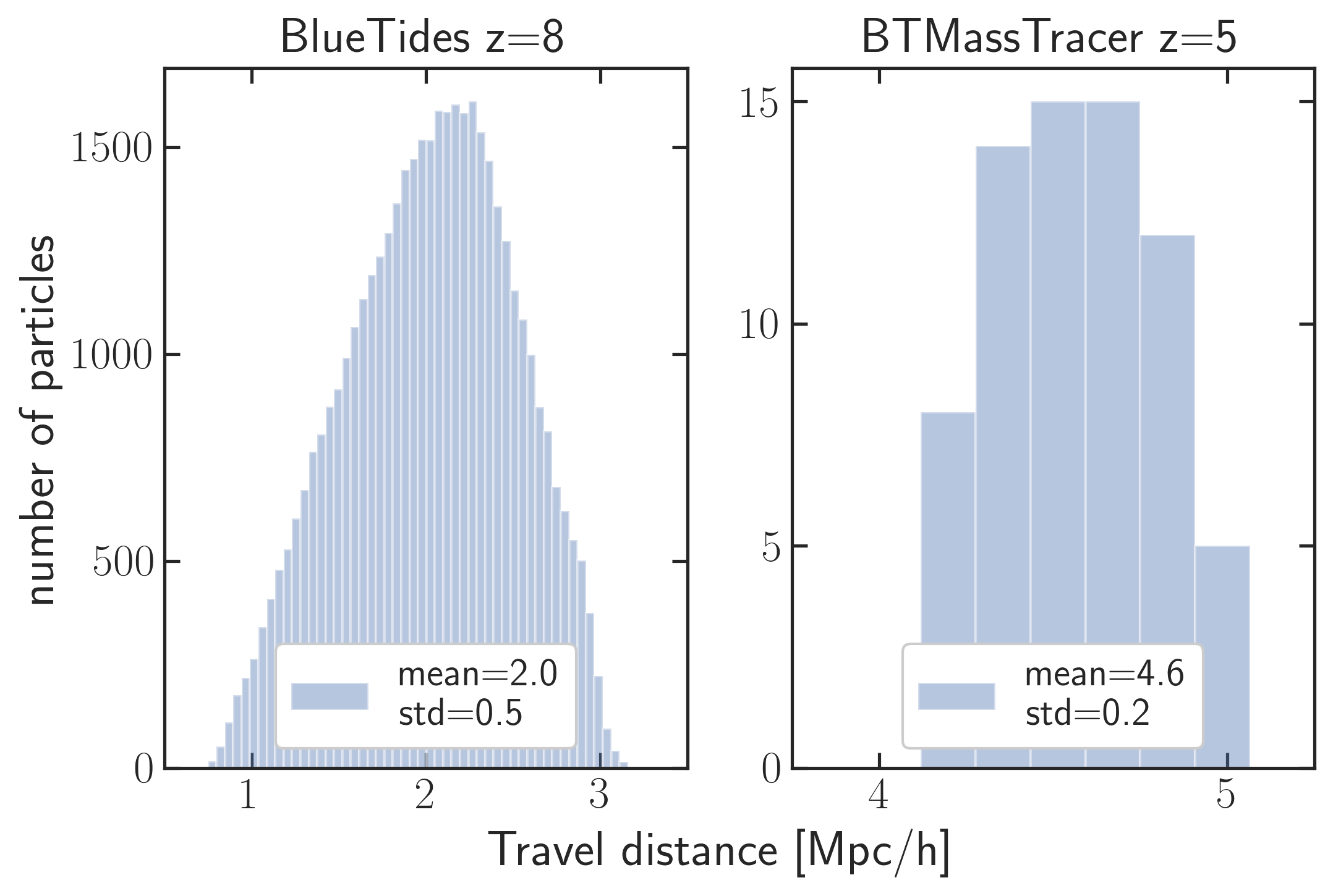}
    \caption{Histogram of the travel distance of all particles in the halo hosting the most massive BH in \BT from $z = 99$ to $z = 8$ and in the same halo in \textsc{BTMassTracer} from $z = 99$ to $z = 5$.}
    \label{fig:travel_dist_bt}
\end{figure}
% Fig ----------------------------------------------

% --------------------------------------------------
We use the massively parallel cosmological smoothed particle hydrodynamic (SPH) simulation software, \textsc{MP-Gadget} \citep{Feng2016}, to run all the simulations in this paper.
Its hydrodynamics solver adopts the new pressure-entropy formulation of SPH \citep{Hopkins2013}.
The main sub-grid models in \textsc{MP-Gadget} are
\begin{itemize}
  \item  star formation based on a multiphase star formation model \citep{Springel2003} with modifications following \citet{Vogelsberger2013},
  \item  gas cooling through radiative processes \citep{Katz1996} and metal cooling \citep{Vogelsberger2014},
  \item  formation of molecular hydrogen and its effects on star formation \citep{Krumholz2011},
  \item  type II supernovae wind feedback \citep{Nelson2015},
  \item  SMBH growth and AGN feedback \citep{DiMatteo2005}.
\end{itemize}

All the new constrained simulations in the paper are run with periodic boundary conditions from $z = 99$ to $z = 6$ (as we are mostly interested in the seed mass and the early growth of SMBHs).
Each simulation contains $2 \times 264^3$ particles in a cube with the box size $L_{\rm box} = 15 \, \hMpc$ (the choice of $L_{\rm box}$ size will be further discussed later in detail).
We adopt the cosmological parameters based on the Nine-Year Wilkinson Microwave Anisotropy Probe Observations \citep{Hinshaw2013}.
All the simulations have the same cosmology and resolution as in \BT, so that we can always use the direct large volume simulations to assess the validity of the new simulations.
Table~\ref{tab:suites} summarizes all the basic parameters of our new runs.
Note that a star particle has a mass of $\Mstar = \frac{1}{4} \, M_{\rm gas}$ and that the gravitational smoothing length $\epsilon$ is the same for all kinds of particles.

Constraining a high-density peak in the initial density field to get a massive halo allows us to study rare objects in a small simulation box rather than in large-volume cosmological hydrodynamic simulations with the box size of a few hundred Mpc on the side.
In order to set a box size for our constrained runs, we need to make sure that the growth history of the halo needs to be well converged.
In particular, we would like to make sure that all the particles that make into the halo and all the way into the central BH are captured.
In particular, here, we want to track the growth of SMBHs at the center and their host galaxies, so we look into how far the particles in the halo hosting the most massive BH in \BT have traveled from $z = 99$ to $z = 8$ in Figure~\ref{fig:travel_dist_bt}.
The mean of the travel distance is $2 \, \hMpc$ with a standard deviation of $0.5 \, \hMpc$, indicating that at least a box size of $L_{\rm box} \geq 2 \, \hMpc$ is necessary to contain the halo up to $z = 8$.
As we will run our new simulations beyond $z = 8$ here we also make use of a dark matter only realization of \BT, the \textsc{BTMassTracer} \citep{Tenneti2018}.
We track all the dark matter particles in the halo down to $z = 5$. We show that the particles typically travel a mean of $4.6 \, \hMpc$ and a standard deviation of $0.2 \, \hMpc$.
This implies an absolute minimum box size of $L_{\rm box} \sim 5 \, \hMpc$.
To be rather conservative and make sure we have the appropriate growth history of the halo and its black holes, we choose a size of $15 \, \hMpc$ and stick to this for all the simulations in the paper.
% --------------------------------------------------

% --------------------------------------------------
\subsection{Constrained versus unconstrained simulations}
\label{subsec:con_vs_uncon}
% --------------------------------------------------

% Fig ----------------------------------------------
\begin{figure*}
    \includegraphics[width=\textwidth]{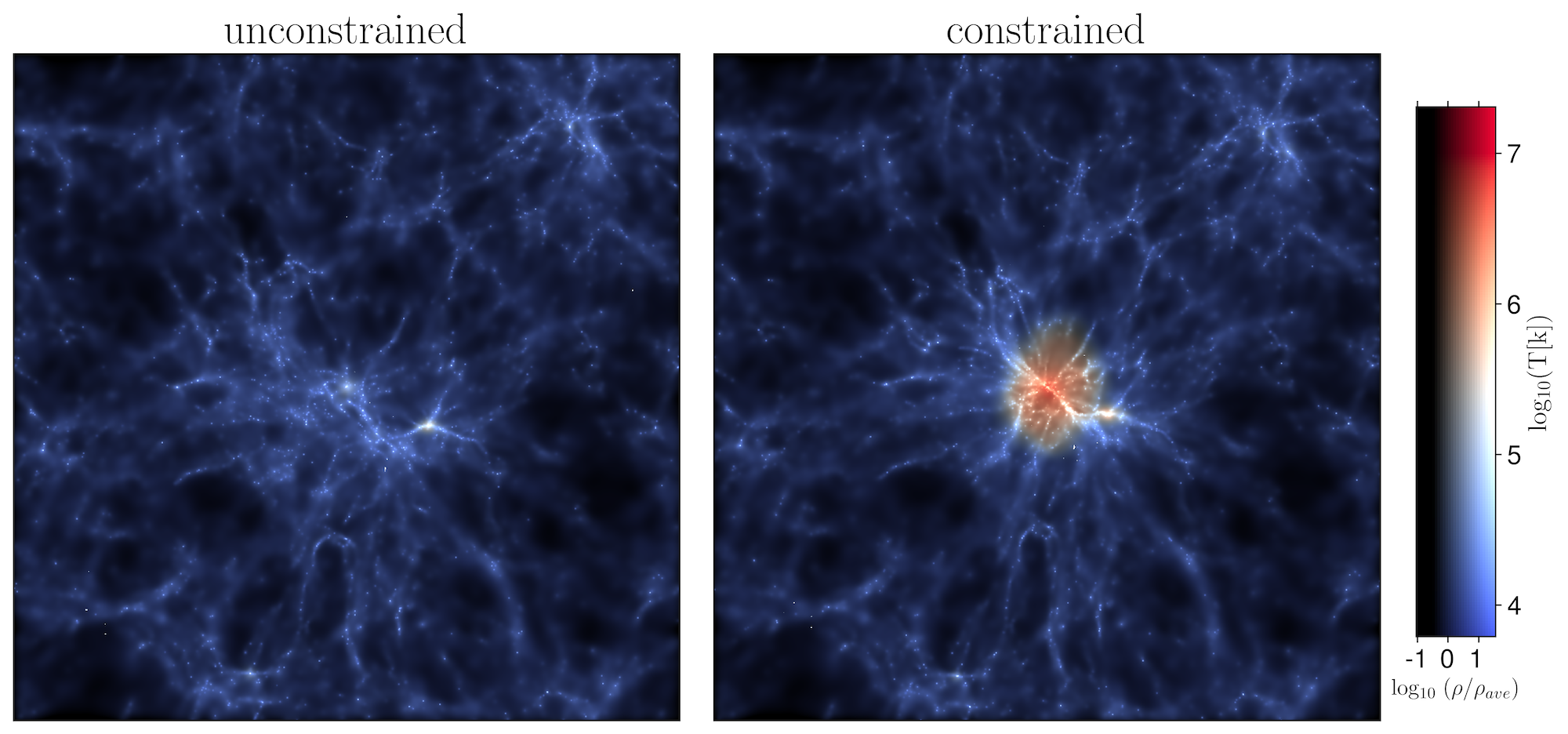}
    \caption{The slices of gas density fields of the unconstrained (left) and constrained (right) simulations at $z = 6$. The gas density field is color-coded by temperature as well. The boxes are $15 \, \hMpc$ per side with a thickness of $5 \, \hMpc$.}
    \label{fig:ic_z=6}
\end{figure*}
% Fig ----------------------------------------------

% Fig ----------------------------------------------
\begin{figure*}
    \includegraphics[width=2\columnwidth]{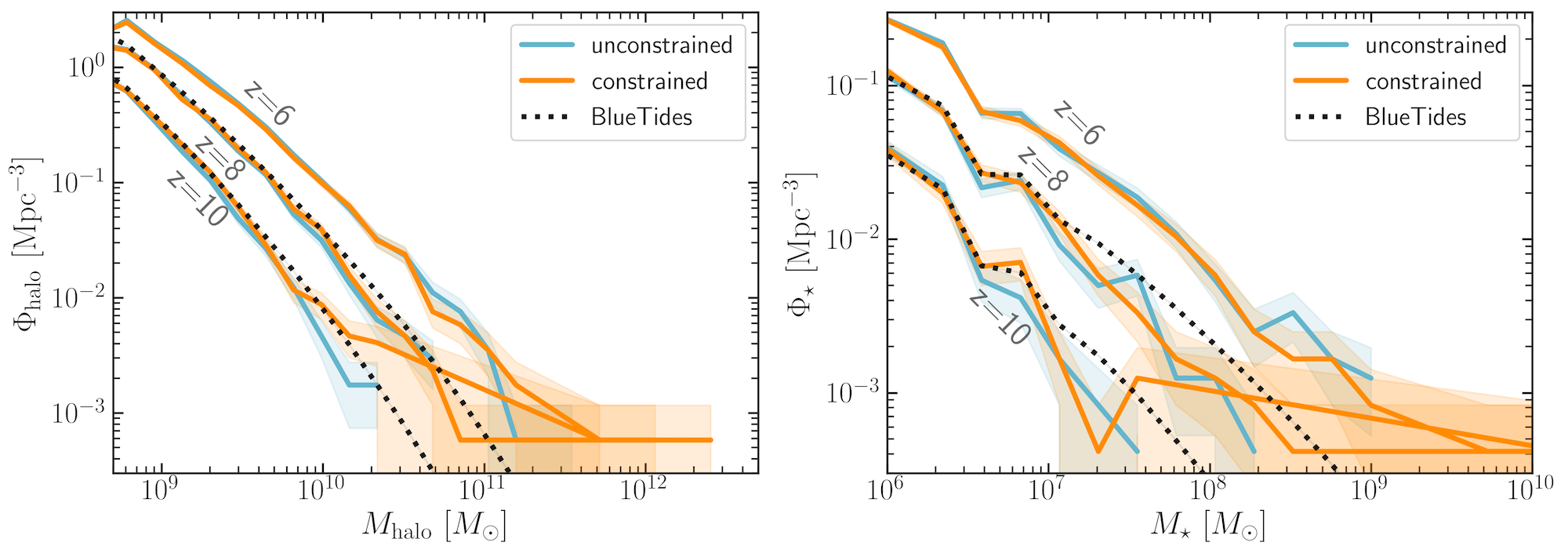}
    \caption{Mass functions in the constrained and unconstrained simulations at $z = 6$, 8, and 10 in comparison with \BT. \textbf{Left}: halo mass functions $\Phi_{\rm halo}$. \textbf{Right}: galaxy stellar mass functions $\Phi_{\rm \star}$.}
    \label{fig:HMFs_STMFs_constraint}
\end{figure*}
% Fig ----------------------------------------------

% Fig ----------------------------------------------
\begin{figure}
    \includegraphics[width=\columnwidth]{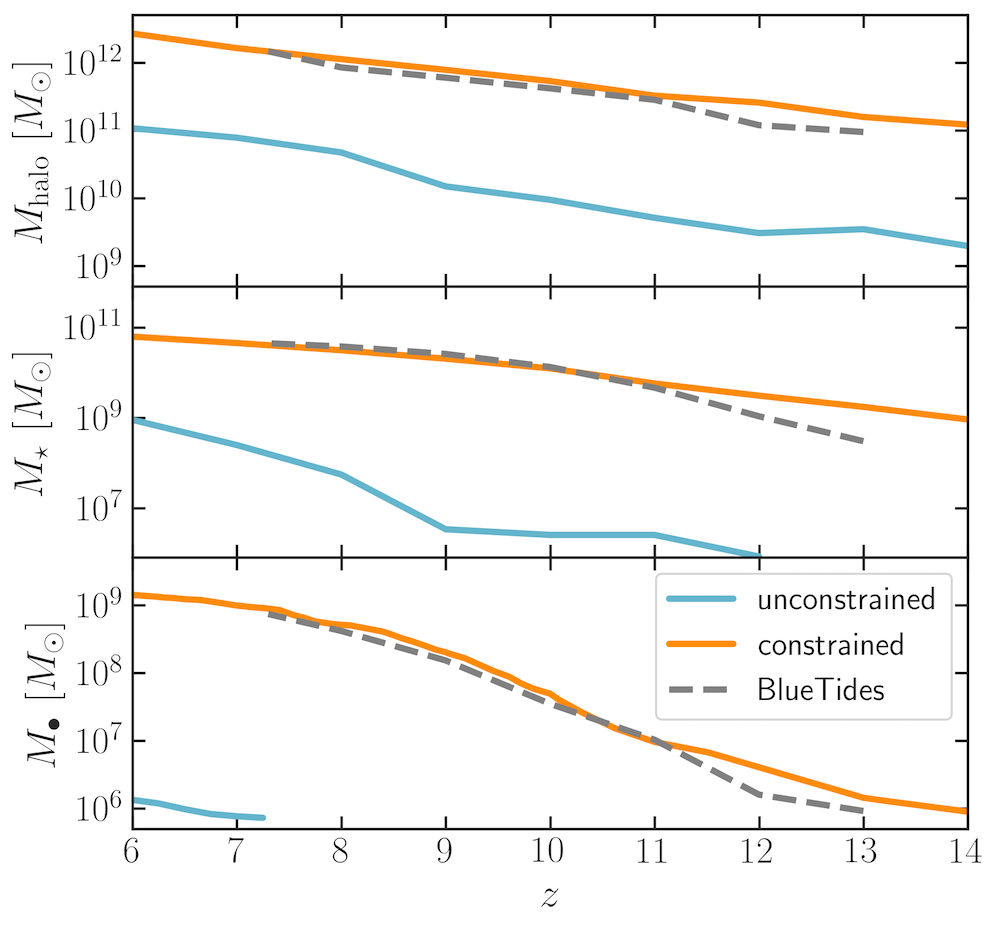}
    \caption{The growth history of the host halo and galaxy ($\Mhalo$ and $\Mstar$) and the most massive BHs ($\Mbh$) in the constrained and unconstrained simulations in comparison with \BT.}
    \label{fig:3massgrowth_constraint}
\end{figure}
% Fig ----------------------------------------------

% --------------------------------------------------
To illustrate the basic features of the constrained simulations, we first run the constrained and unconstrained initial conditions in Figure~\ref{fig:ic_illustration} down to $z = 6$ while keeping all the other simulation parameters.
Figure~\ref{fig:ic_z=6} shows the density fields of the constrained and unconstrained simulations at $z = 6$ color-coded by temperature as well.
As expected, the density around where we put the constrained peak in the constrained simulation is higher than the unconstrained one while the overall structure maintains.
So is the temperature.
Figure~\ref{fig:HMFs_STMFs_constraint} shows the halo and stellar mass functions ($\Phi_{\rm halo}$ and $\Phi_{\rm \star}$) at $z=6$, 8, and 10, compared with \BT.
In particular, there is one halo in the very massive end of these functions in the constrained simulation due to the constrained high-density peak.
Aside from the massive objects, the consistency of both mass functions with each other and with the ones in \BT indicates that the constrained simulation appropriately captures the growth of halo and stellar mass function statistically.

We then investigate the growth history of the most massive BHs ($\Mbh$) and their hosts (halo mass $\Mhalo$ and stellar mass $\Mstar$) in the two simulations compared with that of the \BT simulation in Figure~\ref{fig:3massgrowth_constraint}.
With a proper density peak, the growth history of the three masses in the constrained simulation converges to the ones in \BT (note that a total convergence is not expected as this a new constrained simulation but not a zoom-in simulation).
On the other hand, the halo mass of the unconstrained simulation at $z = 6$ is an order of magnitude less massive than the one in the constrained simulation; the stellar mass is around two orders of magnitude less massive; the BH mass is three orders of magnitude less massive.
% --------------------------------------------------

% --------------------------------------------------
\subsection{Tidal fields of the SMBHs}
\label{subsec:t1}
% --------------------------------------------------

% Fig ----------------------------------------------
\begin{figure*}
    \includegraphics[width=2\columnwidth]{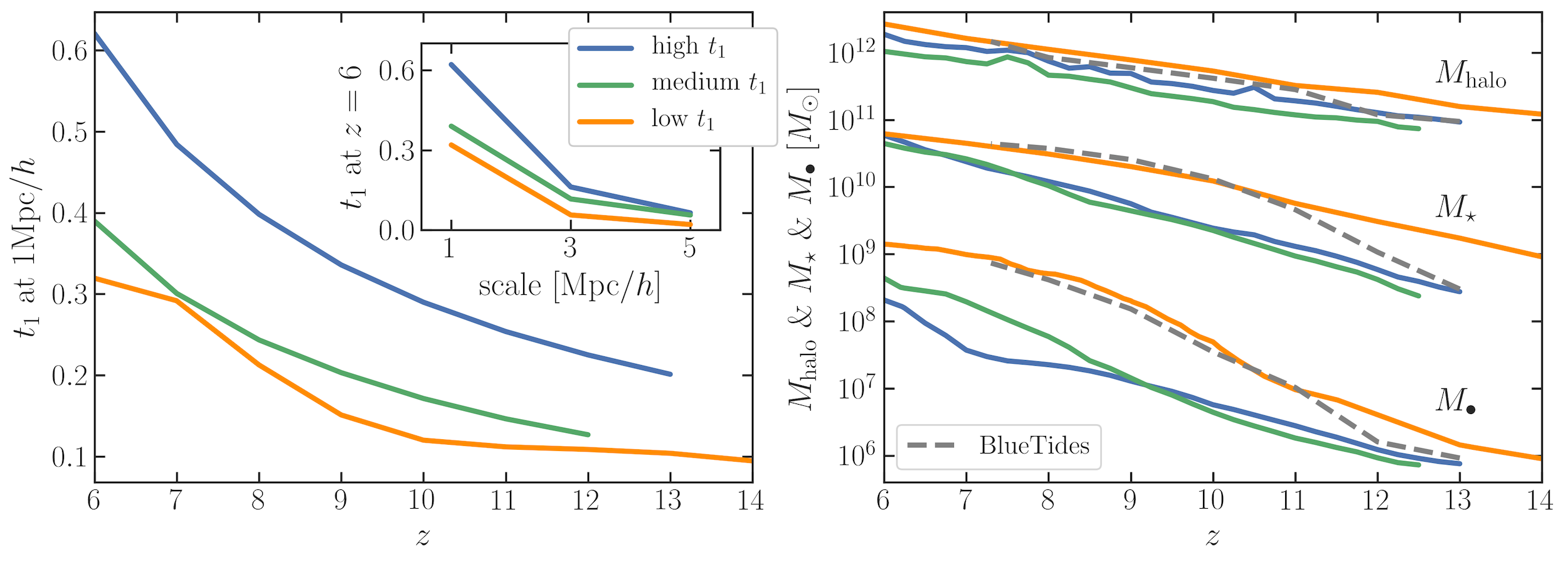}
    \caption{
    \textbf{Left}: tidal field strength $t_1$ measured at the position of the most massive BHs in the three simulations. The inner panel shows $t_1$ measured at different scales at $z = 6$ and the outer panel shows the evolution of $t_1$ measured at $1 \, \hMpc$.
    \textbf{Right}: the growth history of the host halo and galaxy ($\Mhalo$ and $\Mstar$) and the most massive BHs ($\Mbh$) in the three simulations. The grey dashed curves are the quantities of the most massive BH in \BT.}
    \label{fig:t1_all}
\end{figure*}
% Fig ----------------------------------------------

% --------------------------------------------------
While a highly biased region (as in our constrained simulations) is a necessary condition for growing a massive BH, it is not sufficient.
For example in the \BT simulation, only one of the 50 most massive halos of mass similar or greater than the one hosting the most massive BH has a BH more massive than $10^8 \, \Msun$.
As we shall further show, not all of the constrained Gaussian realizations that can grow massive halos guarantee to grow SMBHs in them as well.
Directly related to the density field in the initial conditions, the local tidal field has been identified as the environmental property that is the most strongly correlated to the growth of the first quasars in \citet{DiMatteo2017}.
In these findings, the extreme early growth depends on the early interplay of high gas densities and the tidal field that shapes the mode of accretion in those halos.

The tidal field is characterized by the three eigenvalues ($t_1$, $t_2$, $t_3$) of the local tidal tensor $T_{ij} \equiv S_{ij} - \frac{1}{3} \sum_{i} S_{ii}$, where the strain tensor is the second derivative of the potential, $S_{ij} \equiv \nabla_i \nabla_j \phi$.
According to \citet{Dalal2008}, $S_{ij}$ is calculated in Fourier space as $\hat{S}_{ij} = \frac{k^2}{k_i k_j} \hat{\delta}$.
The three eigenvalues are by definition $t_1 > t_2 > t_3$ and satisfy $t_1 + t_2 + t_3 = 0$ so that $t_1$ is always positive and $t_3$ is negative.
Thus, the tidal field stretches material along with ${\bf t_1}$ and compresses material along with ${\bf t_3}$, where (${\bf t_1}$, ${\bf t_2}$, ${\bf t_3}$) are the corresponding eigenvectors.
To use $t_1$ as the indicator of the local tidal field strength following the standard usage, we calculate $t_1$ numerically using \textsc{nbodykit} \citep{nbodykit}.
We read all the particle data from a snapshot into a mesh object weighted by the particle mass to get $\hat{\delta}$; transform them to Fourier space; apply a kernel of $\frac{k^2}{k_i k_j}$ to get $\hat{S}_{ij}$; transform them back to the real space and evaluate $T_{ij}$ at the position of the SMBH.

To further evaluate the role of the tidal field in the growth of the first massive SMBHs, we generate a number of constrained realizations; select the ones with the minimum, intermediate, and maximum $t_1$ around the density peak as the initial conditions; run them from $z = 99$ to $z = 6$.
The left panel of Figure~\ref{fig:t1_all} shows $t_1$ at the position of the most massive BHs in the three simulations.
$t_1$ measured at $z = 6$ on different scales of 1 -- $5 \, \hMpc$ in the inset suggests that the simulation with a lower or higher $t_1$ is always lower or higher across the scales.
The evolution of $t_1$ measured at the scale of $1 \, \hMpc$ in the main panel shows that a lower or higher $t_1$ environment tends to maintain a lower or higher $t_1$ as time goes.

The right panel of Figure~\ref{fig:t1_all} shows the growth history of the most massive BHs and their hosts in the three simulations.
Several interesting results we find include that, the masses of the halos $\Mhalo$ always differ by a factor less than 10 among the three simulations; the stellar mass $\Mstar$ of the low $t_1$ simulation is around an order of magnitude higher than the others at an earlier stage; the BH masses $\Mbh$ differ by a factor of 10 -- 100.
The three simulations suggest that the tidal field has a larger impact on the growth of SMBHs: a SMBH can grow more or less massive when it is in a lower or higher $t_1$ surrounding environment.
Besides, the growth history of the most massive BH and its host galaxy and halo in the low $t_1$ simulation also converges better to the \BT simulation.

The fact that a lower tidal field environment helps a more massive growth of the SMBHs in our simulations strengthens the findings in \citet{DiMatteo2017} that the local tidal field is strongly correlated to the growth of the first quasars.
Moreover, we utilize the constrained realization that provides the lowest $t_1$ environment as the initial condition for the study of the early growth of SMBHs with different BH seeding parameters in the following sections.
% --------------------------------------------------

\section{Results: different SMBH seeding scenarios}
\label{sec:Results}

% Table --------------------------------------------
\begin{table}
    \centering
    \caption{The sets, names, the BH seed mass $\MbhSeed$, and threshold halo mass $\MfofSeed$ in the simulations.}
    \label{tab:sim_name}
    \begin{tabular}{llcc}
    	\hline
    	Set            & Name               & $\MbhSeed$ [$\hMsun$] & $\MfofSeed$ [$\hMsun$] \\
    	\hline
    	A $\rm ^{a,e}$ & B3H8               & $5 \times 10^{3}$     & $5 \times 10^{8}$ \\
    	A              & B4H9               & $5 \times 10^{4}$     & $5 \times 10^{9}$ \\
    	A, B           & B5H10 $\rm ^{c,d}$ & $5 \times 10^{5}$     & $5 \times 10^{10}$ \\    	
    	B              & B4H10              & $5 \times 10^{4}$     & $5 \times 10^{10}$ \\
    	B $\rm ^b$     & B3H10              & $5 \times 10^{3}$     & $5 \times 10^{10}$ \\
    	\hline
    \end{tabular}\\
    \begin{flushleft}
    $\rm ^a$ Set A contains the simulations with different $\MbhSeed$ and $\MfofSeed$. \\
    $\rm ^b$ Set B contains the simulations with different $\MbhSeed$ only. \\
    $\rm ^c$ B5H10 is the same simulation as the constrained simulation and the low-$t_1$ simulation in Section~\ref{sec:method}. \\
    $\rm ^d$ B5H10 has the same seeding parameters as that of \BT.\\
    $\rm ^e$ \citet{Yu2014MNRAS.440.1865F} has examined the exact pairs of $\MbhSeed$ and $\MfofSeed$ in Set A using zoom-in simulations from the \textsc{MassiveBlack} simulation.
    \end{flushleft}
\end{table}
% Table --------------------------------------------

% Fig ----------------------------------------------
\begin{figure*}
    \includegraphics[width=2\columnwidth]{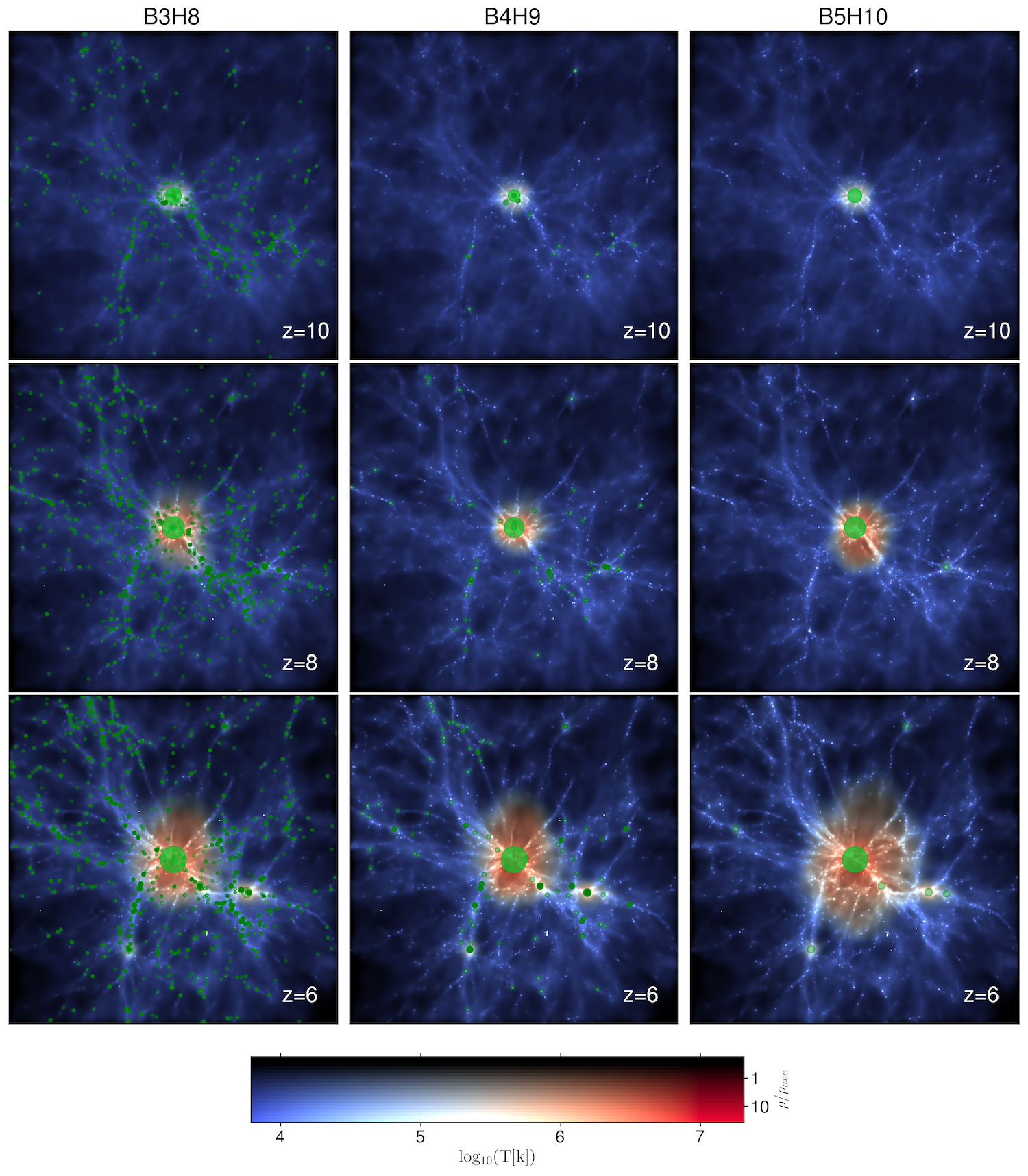}
    \caption{The gas density fields of the simulations B3H8, B4H9, and B5H10 (from the left to the right) at $z = 10$, 8, and 6 (from the top to the bottom). Each of them is centered at the most massive BH with a zoomed-in cube of $6 \, \hMpc$ per side. The gas density fields are color-coded by temperature (blue to red indicating cold to hot respectively, as shown by the color bar at bottom).
    The green marks show the BHs and are sized according to their masses.} 
    \label{fig:run_illustration}
\end{figure*}
% Fig ----------------------------------------------

% Table --------------------------------------------
\begin{table}
    \centering
    \caption{The numbers of BHs in the simulations at different redshifts.}
    \label{tab:number_bh}
    \begin{tabular}{lrrrrr}
    	\hline
    	       & B3H8  & B4H9 & B5H10 & B4H10 & B3H10 \\
    	\hline
    	$z=10$ & 1678  & 36   & 1     & 1     & 1 \\
    	$z=8 $ & 4994  & 135  & 2     & 2     & 2 \\
    	$z=6 $ & 12210 & 488  & 11    & 11    & 11 \\
    	\hline
    \end{tabular}
\end{table}
% Table --------------------------------------------

% Fig ----------------------------------------------
\begin{figure*}
    \includegraphics[width=2\columnwidth]{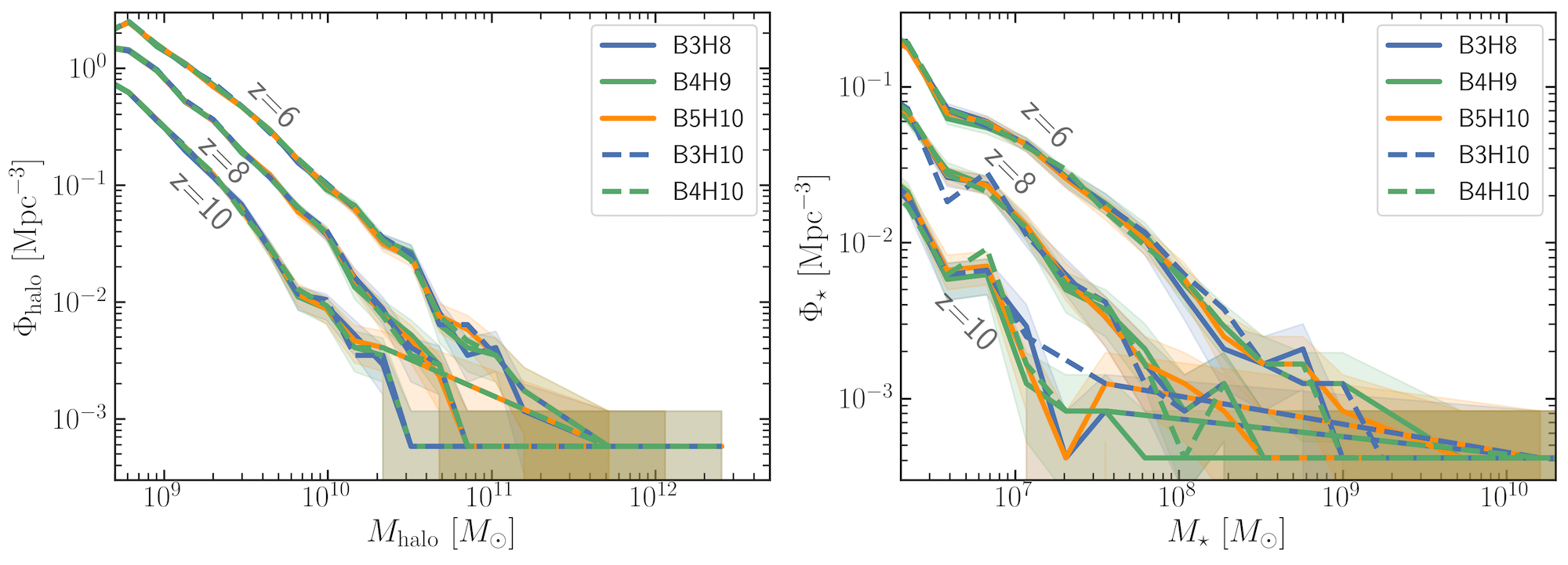}
    \caption{Mass functions of the simulations at $z = 6$, 8, and 10. \textbf{Left}: halo mass functions $\Phi_{\rm halo}$. \textbf{Right}: galaxy stellar mass functions $\Phi_{\rm \star}$.} 
    \label{fig:HMFs_STMFs_15_seedmass}
\end{figure*}
% Fig ----------------------------------------------

% --------------------------------------------------
The main objective of this work is to study the effect of different BH seeding parameters on the early growth of SMBHs using a set of constrained cosmological simulations that statistically reproduce the environments for early growth in the large-volume \BT simulation. 
Here, we conduct two sets of new constrained simulations and investigate the growth history of SMBHs starting from different BH seed masses. In particular, we use three different BH seed masses $\MbhSeed = 5 \times 10^3$, $5 \times 10^4$, and $5 \times 10^5 \, \hMsun$ and perform two sets of simulations. 
\begin{itemize}
    \item  In \textbf{Set A}: we lower the halo mass threshold to $\MfofSeed = 5 \times 10^8$, $5 \times 10^9$, and $5 \times 10^{10} \, \hMsun$ commensurate with keeping the ratio of $\MbhSeed / \MfofSeed$ constant. This is motivated by the physical models for BH seed formation implying that smaller BHs may form earlier in the first, molecular cooling halos which have smaller mass \citep[see, e.g.,][]{Johnsonbromm2007}. Hence in this set, BH seeds with smaller masses (than the canonical $5 \times 10^5 \, \hMsun$) are seeded at earlier times.
    \item  In \textbf{Set B}: we fix the threshold halo mass at $\MfofSeed = 5 \times 10^{10} \, \hMsun$ and study the effect of changing the BH seed mass in a given fixed halo mass. All the BH seeds in this set are seeded at the same time and in the same halos but simply with different BH seed masses. 
\end{itemize}

Table~\ref{tab:sim_name} summarizes the sets of simulations, the adopted naming and their respective BH and halo seeding parameters. We emphasize that all of the simulations have the same constrained initial condition.  
Also, in particular, B5H10 has the same BH seeding parameters as that of \BT with the canonical/reference choice of BH seed mass and halo mass.

To illustrate the results of our simulations, we start by showing the environments of the BHs at $z=6$, 8, and 10 in Figure~\ref{fig:run_illustration}. 
In particular, we show the projected gas density field color-coded by the gas temperature in each of the simulations. 
Each panel is $6 \, \hMpc$ per side with the most massive BH residing at the center. 
The green circles mark out all the BHs, where the size of the circles scales with the BH masses. 
The relatively hot region of gas around the BH results mostly from the effects of AGN feedback.
Here we specifically show the results of the simulations in \textbf{Set A} (B3H8, B4H9, and B5H10). 
We note that the density field/environments of B3H10 and B4H10 in \textbf{Set B} will be similar to those shown for B5H10 except for the gas temperatures in the region around the central BH (which would typically be less affected by AGN feedback and associated heating, as we will discuss later).

Figure~\ref{fig:run_illustration} highlights two major points. 
First, the effects of BH feedback, represented by the heated gas phase (reddish colors) are more prominent as the seed BH grows and as in the case for a single, larger BH seed. 
Second, the BH populations in the simulations B3H8, B4H9, and B5H10 are different. 
This is due to the adopted values for the threshold halo mass in each of the respective simulations. 
In particular, the lower the threshold halo mass is, the more BH seeds are placed in a simulation, resulting in more BHs in B3H8 than in B5H10. 
Table~\ref{tab:number_bh} summarizes the numbers of BHs seeded and growing in each of the simulations.

Figure~\ref{fig:HMFs_STMFs_15_seedmass} shows the halo mass functions $\Phi_{\rm halo}$ and the stellar mass functions $\Phi_{\star}$ in all our simulations (both \textbf{Set A} and \textbf{Set B}).
The consistency of the mass functions among the simulations suggests that the choice of BH seeding parameters does not affect the global halo and galaxy population in significant ways. 
In particular, the lower-mass end is virtually unaffected in both halo and stellar mass functions. 
The high-mass end of the stellar mass function shows some differences. 
As we shall show later, this is a reflection of the different SFR histories (for $z < 10$ in the different seed models which are modulated by different amounts of AGN feedback in different BH seed models.

In the following sections, we aim to explore in more detail the growth histories of most massive BHs and their hosts in the simulations including their masses and mass assembly rates. 
In particular, in Section~\ref{subsec:diffseeddiffhalo} we will show results for the simulations with different BH seed masses and different halo thresholds, B3H8, B4H9, and B5H10 while in
Section~\ref{subsec:diffseedsamehalo} the simulations with different BH seed masses at fixed halo mass, B3H10, B4H10, and B5H10. 
% --------------------------------------------------

% Fig ----------------------------------------------
\begin{figure*}
    \includegraphics[width=2\columnwidth]{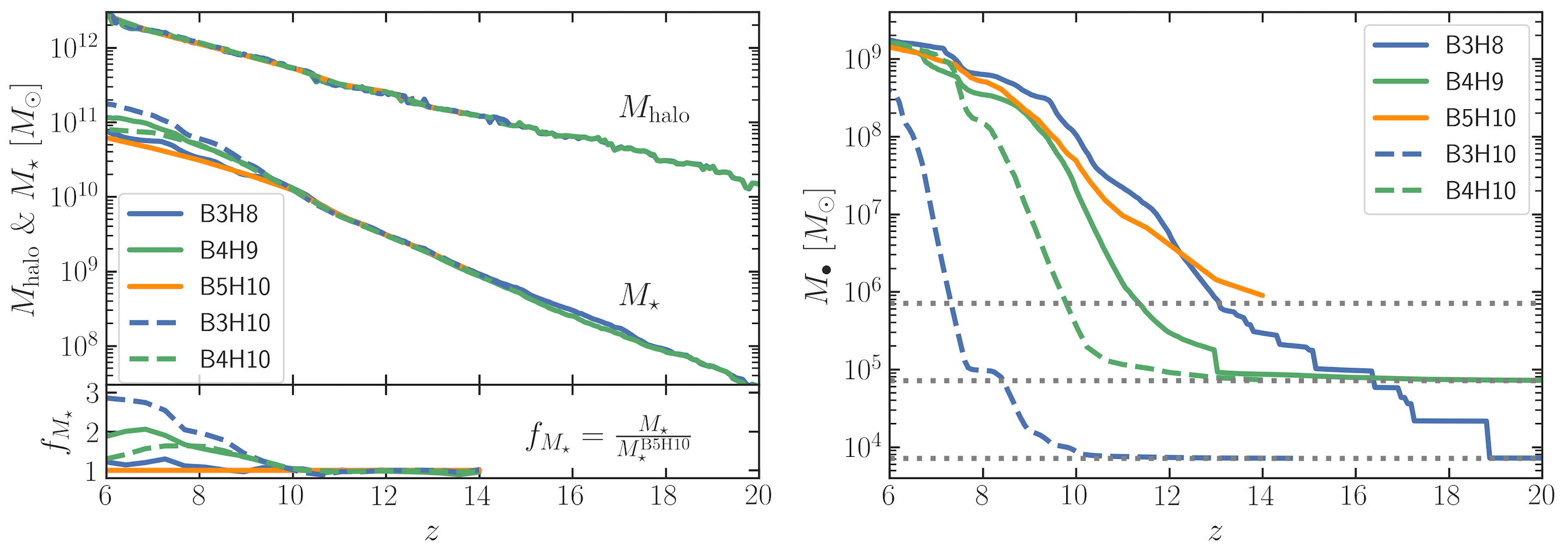}
    \caption{\textbf{Left-top}: the growth history of the host halo and galaxy ($\Mhalo$ and $\Mstar$) of the most massive BHs in the simulations. \textbf{Left-bottom}: the stellar mass ratio $f_{\Mstar}$ between $\Mstar$ in each simulation and in B5H10. \textbf{Right}: the growth history of the most massive BH ($\Mbh$) in each simulation. The horizontal grey dotted lines show the BH seed masses.} 
    \label{fig:Mbh_z}
\end{figure*}
% Fig ----------------------------------------------

% Fig ----------------------------------------------
\begin{figure*}
    \includegraphics[width=2\columnwidth]{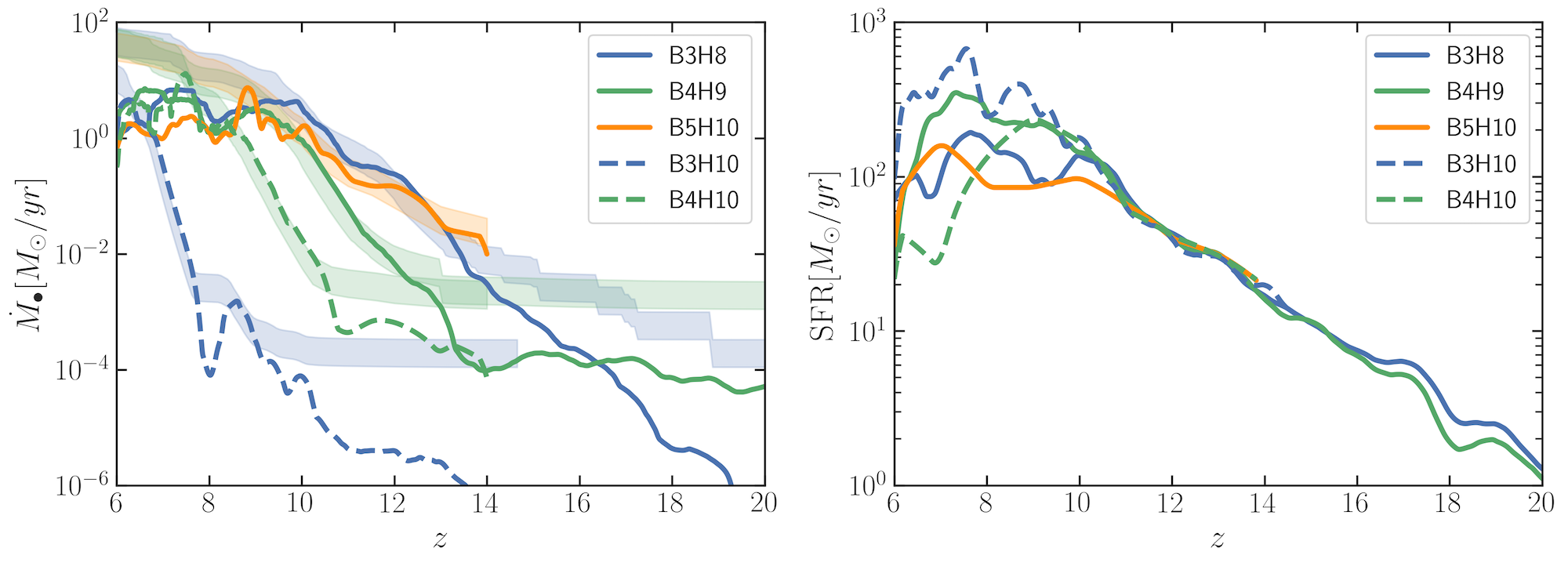}
        \caption{\textbf{Left}: the BH accretion rate ($\dot{M}_{\bullet}$) of the most massive BHs in the simulations. The shady regions show the Eddington rates of the BHs. \textbf{Right}: the star formation rate (SFR) of the most massive BHs in the simulations.} 
        \label{fig:bhacc_sfr_seedmass}
\end{figure*}
% Fig ----------------------------------------------

% --------------------------------------------------
\subsection{Set A: different BH seed masses and halo mass thresholds}
\label{subsec:diffseeddiffhalo}
% --------------------------------------------------

% --------------------------------------------------
Here we describe the results from the simulations B3H8, B4H9, and B5H10. Those are the ones in which halo thresholds for BH seeding are adjusted such that the ratio $\MbhSeed / \MfofSeed$ is fixed. 
The left-top panel of Figure~\ref{fig:Mbh_z} shows the growth history of host halo and galaxy ($\Mhalo$ and $\Mstar$) for the most massive BHs in the simulations. 
The halos show the same growth history among the simulations and their masses at $z = 6$ are $3 \times 10^{12} \, \hMsun$. 
The galaxy stellar mass also have a similar growth history except for B4H9 in which the host stellar mass ends up being larger than the others after $z = 8$ but less than a factor of two different at $z = 6$ according to the stellar mass ratio in the left-bottom panel. 
The stellar mass ratio $f_{\Mstar} = \frac{\Mstar}{\Mstar^{\rm B5H10}}$ compares the galaxy mass in each of the simulation to $\Mstar$ in B5H10. 
At $z = 6$, $\Mstar = 10^{11} \, \hMsun$ for B3H8 and B5H10 and $\Mstar = 6 \times 10^{10} \, \hMsun$ for B4H9. 
Our simulations suggest that the choice of BH seeding parameters does not affect the growth of the hosts for more than a factor of two.

Of our interest, the solid curves in the right panel of Figure~\ref{fig:Mbh_z} show the growth history of the most massive BHs in the simulations B3H8, B4H9, and B5H10.  
As apparent in the figure, the BHs are seeded with different masses according to $\MbhSeed$ and at different times according to $\MfofSeed$; that is, a smaller BH seed emerges in a lower-mass halo at an earlier time than a corresponding higher-mass one. 
For example, when a BH seed of $5 \times 10^3 \, \hMsun$ is placed in halos of mass $5 \times 10^8 \, \hMsun$ it can be seeded at $z > 20$. 
As halos of this mass are not so rare at high redshifts, a lot of BH seeds emerge rather than just one seed does as in the case of $\MbhSeed = 5 \times 10^5 \, \hMsun$. 
The most massive BHs in the simulations start to converge in mass at $z \sim 8$ and reach a mass of $2 \times 10^9 \, \hMsun$ by $z = 6$ even though the seed population and the total number of BHs are different. 
This suggests that the choice of different pairs of BH seed mass and threshold halo mass does not significantly affect the growth of the most massive BHs, at least in the expected range of $\MbhSeed = 10^3$ -- $10^6 \, \Msun$. 
However, the early growth of SMBHs in the three simulations can be faster or slower at $z > 10$; that is, the small and large BH seeds start more massively while the intermediate one remains at its seed mass the longest but catch up drastically once it starts growing. 
Moreover, the discrete jumps in the growth history of B3H8 at early times indicate a lot of mergers occur even at such high redshifts compared to the others.

Figure~\ref{fig:bhacc_sfr_seedmass} shows the evolution of BH accretion rates $\Dot{M}_{\bullet}$ of the most massive BHs in the simulations. 
The solid curves show the $\Dot{M}_{\bullet}$ and the shaded bands indicate the regime between the Eddington rate and two times the Eddington rate as the upper limit for $\Dot{M}_{\bullet}$ in the simulations. 
We find that $\Dot{M}_{\bullet}$ has the same overall evolution: starting with an initial low accretion phase, followed by a close to exponential Eddington growth, and ending with a final quenched feedback-regulated phase. 
It is noticeable that $\Dot{M}_{\bullet}$ gradually converges at $z > 10$, the time when the three SMBHs enter the feedback-regulated phase where their growth saturates and $\Mbh$ starts converging. 
However, in the early phases during $10 < z < 13$, the BH accretion rates have different trajectories as the BHs experiencing exponential growth but constrained by the upper limit in the simulations.

The right panel of Figure~\ref{fig:bhacc_sfr_seedmass} shows the star formation history of the host galaxies for the simulations. 
The evolution of the star formation rates (SFRs) appears similar at $z > 10$ but diverges in the later phase because the AGN feedback starts to regulate the star formation rate by coupling significant energy to the star forming gas.
Therefore, $\Dot{M}_{\bullet}$ and SFR start to couple with each other after a significant BH growth phase ($z < 10$) though the physical scales of the two quantities are quite different (SFR is in the galactic scale of tens of $\hkpc$, whereas $\Dot{M}_{\bullet}$ is determined by local gas density at the scale of $\hkpc$).

As a further comparison, \citet{Yu2014MNRAS.440.1865F} has examined the exact pairs of $\MbhSeed$ and $\MfofSeed$ in \textbf{Set A} (see Table~\ref{tab:sim_name}) using zoom-in simulations. 
With the zoom-in technique, they re-simulated a high-redshift ($z > 5.5$) halo hosting a $10^9 \, \Msun$ BH from the $\rm \sim Gpc$ volume, \textsc{MassiveBlack} cosmological hydrodynamic simulation. 
They reported that regardless of the BH seed mass, the BH masses converged to $\Mbh = 10^9 \, \Msun$ at $z = 6$, the BHs underwent a similar history of BH accretion, and the evolution of SFRs was the same at the earlier times before AGN feedback starts to regulate the SFR. 
It is interesting that both their findings and ours are fully consistent with completely different methods, the zoomed-in simulation and the constrained simulation. 
This strengthens the conclusion that the choice of BH seed mass - when keeping the ratio $\MbhSeed / \MfofSeed$ fixed - does not affect the growth of SMBHs significantly.
% --------------------------------------------------

% --------------------------------------------------
\subsection{Set B: different BH seed masses at fixed host halo mass }
\label{subsec:diffseedsamehalo}
% --------------------------------------------------
We then move on to investigate the growth history of host halo and galaxy ($\Mhalo$ and $\Mstar$) of the most massive BHs in the simulations B3H10, B4H10, and B5H10 in Figure~\ref{fig:Mbh_z}. 
The halos show the same growth history among the simulations and their masses at $z = 6$ are $3 \times 10^{12} \, \hMsun$. 
On the other hand, the galaxies seem to have a very similar growth history at $z > 10$ but then their masses start to differ.
According to the stellar mass ratio $f_{\Mstar}$, the galaxy in B3H10 is three times more massive than the one in B5H10 at $z = 6$. 
This can be inferred through the evolution of SFR in Figure~\ref{fig:bhacc_sfr_seedmass}; that is, the galaxy in B3H10 undergoes a rather bursty star formation history at $z < 10$ with variations in SFR up to an order of magnitude compared to the others. 
The reason is that the BH mass in B3H10 is smaller than the others and therefore the corresponding AGN feedback is not strong enough to bring sufficient suppression on the local star formation at $6 < z < 10$. 
Our simulations suggest that the choice of BH seed mass does not affect the growth of halo by $z = 6$; does not affect the growth of galaxy by $z = 10$; do affect the growth of galaxy after $z = 10$ but less than a factor of three by $z = 6$. 
There is another noticeable trend that a galaxy grows more when the BH seed mass is smaller.

The growth of the most massive BHs in Figure~\ref{fig:Mbh_z} starts with different masses and then converges to $\sim 10^9 \, \Msun$ by $z = 6$ except for the one in B3H10, which is less than an order of magnitude difference. 
The evolution of BH accretion rates of the BHs in Figure~\ref{fig:bhacc_sfr_seedmass} shows that the BH in B3H10 is still experiencing Eddington exponential growth at $z < 8$. 
This indicating that the BH in B3H10 is still catching up in mass and will probably converge to the others at later times. 
Our simulation results hint that the growth of the most massive BHs will converge at the later times regardless of the choice of the SMBH seeding parameters in cosmological simulations. 
% --------------------------------------------------

% --------------------------------------------------
\subsection{BH-BH mergers}
\label{subsec:mergers}
% --------------------------------------------------

% Fig ----------------------------------------------
\begin{figure*}
    \includegraphics[width=2\columnwidth]{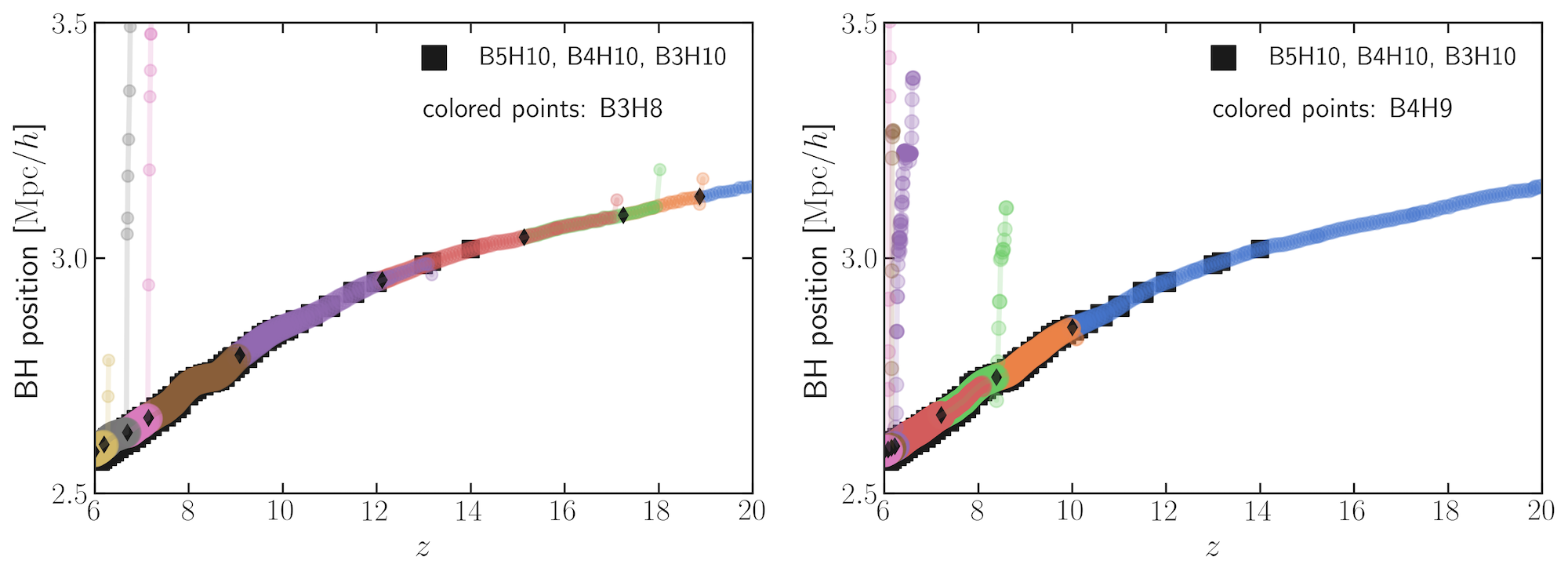}
    \caption{Positions of the most massive BHs in simulations B3H8 (left) and B4H9 (right) compared with B5H10 (and the others). The black diamonds mark the mergers the two most massive BHs experience. The size and color of the data points illustrate the mass of BHs and the ID of BH particles.} 
    \label{fig:mergers_15Mpc_pos}
\end{figure*}
% Fig ----------------------------------------------

% --------------------------------------------------
Figure~\ref{fig:run_illustration} and Table~\ref{tab:number_bh} indicate that there is a major difference in the BH populations of the simulations B3H8, B4H9 and B5H10. 
Particularly in B3H8, there is a vast BH population in the environment of the most massive BH since we also adjust the minimum halo, implying that BH mergers are more likely to happen in the early times. 
The step-like feature in the BH mass assembly history in Figure~\ref{fig:Mbh_z} then infers that mergers occur. 
These pieces of evidence motivate the following investigation of BH mergers in the simulations. 
In \textsc{MP-Gadget}, a BH-BH merger occurs when the distance between two BHs is smaller than their SPH smoothing kernel and the relative velocity of the two BHs is smaller than $\frac{1}{2} \, c_{\rm s}$, where $c_{\rm s}$ is the local sound speed of the gas. 
Convection is also applied to the BH dynamics by repositioning the BH particle to the local minimum potential at every time step.

To investigate the BH merger history of the simulations B3H8, B4H9, and B5H10 (as well as B3H10 and B4H10), we plot the merger tree of BH projected positions in Figure~\ref{fig:mergers_15Mpc_pos}. 
The left and right panels show the position of the most massive BHs in B3H8 and B4H9 respectively compared with B5H10. 
The size of the data points scales with the BH masses; the data points are color-coded by the ID of BH particles; the black diamonds mark where and when the BH mergers occur. 
By $z = 6$, there are eight and six BH mergers that happen in B3H8 and B4H9 respectively whereas there is no merger in B5H10 despite mergers likely to occur below this redshift as the closest BH below 100~kpc distance from the most massive one. 
Besides, B3H10 and B4H10 have no merger as well as B5H10 since they contain the same number and position of BHs but with smaller BH masses.

At $z > 12$, interestingly, four mergers happen in B3H8 whereas none in the others. 
This explains why the SMBH in B3H8 grows faster than the others during the earlier phase and further implies that mergers dominate the early growth of SMBHs in small BH seeding scenarios. 
Our simulations suggest that if less-massive BH seeds are more common, SMBHs can still grow via mergers at the early times even though they may be expected to grow slower. 
In other words, a different BH merger history results in a different growth of the SMBHs particularly at the early times. 
Despite the fact that the high halo occupation fraction of SMBHs in cosmological simulations will increase the number of BH-BH mergers since these simulations are implemented with simple merger models without considering the BH dynamics that could make BH-BH mergers more difficult between low-mass BHs, it is still interesting that different seeding scenarios are expected to produce different BH populations and associated merger rates that can discriminate the different scenarios at early times while the final BH mass converges to a similar value.

In contrast, at $6 < z < 10$, SMBHs seeded with a mass of $10^3$ -- $10^4 \, \Msun$ undergo a few BH mergers whereas the one in the largest seed models ($\sim 10^5 \, \Msun$) does not experience any merger until $z < 6$. 
These different predictions of the merger history for the first massive BHs constitute an interesting prospect for constraining BH seed masses or models for the first quasars that will become within reach with the planned \textit{LISA} mission \citep{2017arXiv170200786A}. 
% --------------------------------------------------

\section{Conclusion}
\label{sec:Conclusions}
In this paper, we have investigated new constrained cosmological simulations designed to reproduce the environments and large-scale structures relevant for the growth of the first quasars at $z \geq 6$. 
In particular, we have focused on the effects of different choices of BH seeding scenarios (different parameters in the SMBH sub-grid model) on the growth of SMBHs at the early times. 
Employing the technique of constrained Gaussian realizations \citep{Hoffman1991, vandeWeygaert1996}, we have reconstructed the initial conditions to reproduce the large-scale structure and the local environment of the most massive BH in the \BT simulation. 
\BT has been the only cosmological hydrodynamic simulation that directly predicted the rare-observed first quasars \citep{DiMatteo2017, Ni2018, Tenneti2018} thanks to its sufficiently high resolution and large volume. 
The first quasars are extremely rare such that there were only four SMBHs with mass $\sim 10^9 \, \Msun$ by $z = 7$ in \BT with $L_{\rm box} = 400 \, \hMpc$.

We have compared the new constrained simulations with the \BT simulation to validate this method by running the constrained initial conditions forward in time until $z = 6$. 
Our new simulations in boxes of $15 \, \hMpc$ on a side have successfully recovered the evolution of the large-scale structure, the mass functions, as well as the growth history of the most massive BHs and their hosts at the high redshifts of interests. 
At $z = 8$, the most massive BH and its hosts had a halo mass of $\sim 10^{12} \, \Msun$; a stellar mass of $\sim 4 \times 10^{10} \, \Msun$; a BH mass of $\sim 4 \times 10^8 \, \Msun$. 
This is consistent with \BT within a factor of 1.5 in mass while keeping the resolution. 
More importantly, the demand on computational resources has decreased significantly by a factor of $(400 / 15)^3 \sim 20000$.

By running a set of different realizations such that each of them has a different local tidal field, we have further shown that a low-tidal field environment is crucial for the growth of the earliest and most massive SMBHs.
This is consistent with the finding from \BT in \citet{DiMatteo2017}. 
For our highest tidal field realization, the mass of the most massive BH was only $\sim 2 \times 10^7 \, \Msun$ at $z = 7$ which was two orders of magnitude lower than the SMBH in the lowest tidal field realization. 
Among the simulations, the SMBH in the lowest tidal field environment had a mass an order of magnitude more than the one in the highest tidal field environment at $z = 6$.

After selecting the initial conditions that best recovered the original quasar environment in \BT, we have run other simulations to investigate the effects of the choice of BH seeding parameters on the growth of these first massive objects. 
In \BT simulation, the BH seed mass has been chosen to be $5 \times 10^5 \, \hMsun$, which is at the high end of the predicted mass for SMBH seeds in theories. 
With the same constrained initial conditions, we have conducted two sets of simulations with different SMBH seed masses $\MbhSeed = 5 \times 10^3$, $5 \times 10^4$, and $5 \times 10^5 \, \hMsun$.
Set A with different threshold halo masses such that the ratio $\MbhSeed / \MfofSeed$ is fixed, while set B has a fixed halo threshold. 
Our simulations have suggested that the final mass of the SMBH is insensitive to the initial seed mass regardless of the choice of BH seeding parameters; the mass of SMBH in our constrained simulations has converged to $\sim 10^{9} \, \Msun$ at $z = 6$. 
In the early times at $z > 10$, the growth of SMBHs varies among the simulations with different seeding scenarios; less massive seed models tend to grow slower initially unless they are seeded in more common but less massive halos so that they can merge frequently.
A significant fraction of the early growth occurs in this mode in a low mass seed scenario in set A, effectively allowing the SMBH growth to catch up with that of a more massive seed. 
There were four SMBH mergers at $z \gtrsim 12$ for the most massive SMBH with the lowest seed mass while no mergers happened for the other two runs, suggesting that the smallest seed grows faster at earlier times when seeded in less massive halos.

The significant differences in the early merger rates provide an interesting discriminating feature for small versus large BH seed models at the early time. 
The space-based gravitational wave telescope \textit{LISA} will open up new investigations into the dynamical processes involving SMBHs and new exciting prospects for tracing the origin, merger history of SMBHs across cosmic ages.

\section*{Acknowledgements}
We acknowledge the funding from NSF ACI-1614853, NSF AST-1517593, NSF AST-1616168, NASA ATP 80NSSC18K1015 and NASA ATP 17-0123.
The \BT simulation is run on the BlueWaters facility at the National Center for Supercomputing Applications.
Some of the simulations in this work are carried out on the Stampede2 supercomputing cluster.
The authors acknowledge the Texas Advanced Computing Center (TACC) at the University of Texas at Austin for providing HPC resources that have contributed to the research results reported within this paper. URL: \href{http://www.tacc.utexas.edu}{http://www.tacc.utexas.edu}.

This work has made use of the following software as well:
\acknowledgeSoftware{numpy},
\acknowledgeSoftware{scipy},
\acknowledgeSoftware{matplotlib},
\acknowledgeSoftware{seaborn},
\acknowledgeSoftware{gaepsi2},
\acknowledgeSoftware{nbodykit}.

%%%%%%%%%%%%%%%%%%%%%%%%%%%%%%%%%%%%%%%%%%%%%%%%%%


\bibliographystyle{mnras}
\bibliography{bib.bib}

%%%%%%%%%%%%%%%%%%%%%%%%%%%%%%%%%%%%%%%%%%%%%%%%%%

%%%%%%%%%%%%%%%%% APPENDICES %%%%%%%%%%%%%%%%%%%%%

%\appendix

%%%%%%%%%%%%%%%%%%%%%%%%%%%%%%%%%%%%%%%%%%%%%%%%%%

% Don't change these lines
\bsp	% typesetting comment
\label{lastpage}
\end{document}